\documentclass[runningheads]{llncs}
\usepackage[utf8]{inputenc}
\usepackage[T1]{fontenc}
\usepackage[usenames,dvipsnames,svgnames,x11names,table]{xcolor}
\usepackage{cite}
\usepackage{amsmath,amssymb,amsfonts}
\usepackage{graphicx}
\usepackage{textcomp}
\usepackage{listings}
\usepackage{url}
\usepackage{pifont}
\usepackage[colorlinks=true,linkcolor=blue,urlcolor=blue,citecolor=blue,bookmarks=false]{hyperref}\usepackage[abbreviations]{foreign}
\usepackage{booktabs} \def\BibTeX{{\rm B\kern-.05em{\sc i\kern-.025em b}\kern-.08em
    T\kern-.1667em\lower.7ex\hbox{E}\kern-.125emX}}
\usepackage{multirow}
\usepackage[abbreviations]{foreign}
\usepackage[inline]{enumitem}
\usepackage{cleveref}\usepackage{subcaption}
\usepackage[subtle]{savetrees}
\usepackage[margin=0pt,skip=6pt,belowskip=0pt,font={small,stretch=0.9},labelfont=bf]{caption}
\usepackage{tabularx, makecell, booktabs}
\usepackage{tikz}
\usepackage{mathtools}
\usepackage{algorithm}
\usepackage{algpseudocode}
\usepackage{siunitx}
\usepackage[export]{adjustbox}
\usepackage{eso-pic}
\usepackage[most]{tcolorbox}
\usepackage{orcidlink}
\definecolor{prioritycolor}{HTML}{969bce}
\newcommand*{\priority}[1]{\raisebox{-1pt}{\begin{tikzpicture}[scale=0.12]\draw (0,0) circle (1);
  \fill[fill opacity=1,fill=prioritycolor] (0,0) -- (90:1) arc (90:90-#1*3.6:1) -- cycle;
  \end{tikzpicture}}}

\newcommand{\compfull}{\priority{100}}
\newcommand{\comppart}{\priority{50}}
\newcommand{\compnone}{\priority{0}}

\makeatletter
\renewcommand\section{\@startsection{section}{1}{\z@}{-8\p@ \@plus -4\p@ \@minus -4\p@}{6\p@ \@plus 4\p@ \@minus 4\p@}{\normalfont\large\bfseries\boldmath
                        \rightskip=\z@ \@plus 8em\pretolerance=10000 }}
\renewcommand\subsection{\@startsection{subsection}{2}{\z@}{-8\p@ \@plus -4\p@ \@minus -4\p@}{6\p@ \@plus 4\p@ \@minus 4\p@}{\normalfont\normalsize\bfseries\boldmath
                        \rightskip=\z@ \@plus 8em\pretolerance=10000 }}
\renewcommand\subsubsection{\@startsection{subsubsection}{3}{\z@}{-6\p@ \@plus -4\p@ \@minus -4\p@}{-0.5em \@plus -0.22em \@minus -0.1em}{\normalfont\normalsize\bfseries\boldmath}}
\renewcommand\paragraph{\@startsection{paragraph}{4}{\z@}{-4\p@ \@plus -4\p@ \@minus -4\p@}{-0.5em \@plus -0.22em \@minus -0.1em}{\normalfont\small\itshape}}\makeatother

\newcolumntype{R}{>{\raggedleft\arraybackslash}X}
\newcolumntype{Y}{>{\centering\arraybackslash}X}
\newcolumntype{P}[1]{>{\raggedleft\arraybackslash}p{#1}}
\newcolumntype{q}[1]{>{\centering\arraybackslash\hspace{0pt}}p{#1}}

\newboolean{showcomments}
\setboolean{showcomments}{true}
\ifthenelse{\boolean{showcomments}}
{ \newcommand{\mynote}[3]{
   \fbox{\bfseries\sffamily\scriptsize#1}
   {\small$\blacktriangleright$\textsf{\emph{\color{#3}{#2}}}$\blacktriangleleft$}}}
{ \newcommand{\mynote}[3]{}}

\newcommand{\sys}{\textsc{BlindexTEE}\xspace}

\makeatletter
\begin{document}
\title{\sys: A Blind Index Approach towards TEE-supported End-to-end Encrypted DBMS}
\titlerunning{\sys}
\author{Louis Vialar\orcidlink{0009-0006-9875-7635} \and
Jämes Ménétrey\orcidlink{0000-0003-2470-2827} \and
Valerio Schiavoni\orcidlink{0000-0003-1493-6603} \and
Pascal Felber\orcidlink{0000-0003-1574-6721}}
\authorrunning{L. Vialar \and J. Ménétrey \and V. Schiavoni \and P. Felber}
\institute{University of Neuchâtel, Neuchâtel, Switzerland\\
\email{first.last@unine.ch}}
\maketitle              \begin{abstract}
    Using cloud-based applications comes with privacy implications, as the end-user looses control over their data.
    While encrypting all data on the client is possible, it largely reduces the usefulness of database management systems (DBMS) that are typically built to efficiently query large quantities of data.
    We present \sys, a new component that sits between the application business-logic and the database.
    \sys is shielded from malicious users or compromised environments by executing inside an SEV-SNP confidential VM, AMD's trusted execution environment (TEE).
    \sys is in charge of end-to-end encryption of user data while preserving the ability of the DBMS to efficiently filter data.
    By decrypting and re-encrypting data, it builds \emph{blind indices}, used later on to efficiently query the DBMS.
    We demonstrate the practicality of \sys with MySQL in several micro- and macro-benchmarks, achieving overheads between 36.1\% and 462\% over direct database access depending on the usage scenario.

\end{abstract}
\definecolor{yellowPaper}{HTML}{fff8ae}
\AddToShipoutPictureFG*{\AtTextUpperLeft{
    \adjustbox{raise=59pt,center}{\begin{tcolorbox}[width=\textwidth,enhanced,colback=yellowPaper,frame hidden,sharp corners]
        \scriptsize
        \copyright~2024\ Springer Nature Switzerland AG.
        This preprint has not undergone peer review (when applicable) or any
        post-submission improvements or corrections. The Version of Record of
        this contribution is published in Proceedings of the 26th International Symposium on Stabilization, Safety, and Security of Distributed Systems, and is
        available online at \href{https://doi.org/10.1007/978-3-031-74498-3_19}{https://doi.org/10.1007/978-3-031-74498-3\_19}
     \end{tcolorbox}
  }}}

\section{Introduction}
\label{sec:intro}

The high convenience of the modern web-based application, accessible everywhere from any device, comes with important privacy downsides.
Once data is off-loaded to third-party service providers, one never knows its future usage. Many solutions exist to protect data stored in remote untrusted database management systems (DBMS)~\cite{fuller_sok_2017,popa_cryptdb_2011,DBLP:conf/sigmod/AntonopoulosASE20,DBLP:journals/popets/VinayagamurthyG19,DBLP:conf/nana/WangLSMWYSLZD17}.
These solutions protect data stored by the service provider from malicious database administrators, but don't protect user data from the service provider itself.

We present \sys, a novel approach for database encryption.
In a nutshell, data is encrypted in such a way that only the data owners (end users of the system) can access it, while preserving the possibility to retrieve the data efficiently from the database system.
\sys is a database proxy that transparently sits between the database client and database server.
It handles on-the-fly encryption and decryption of data for confidentiality and makes use of \emph{blind indices}~\cite{noauthor_ciphersweet_nodate} for efficient retrieval of encrypted data in the database.
A blind index is a kind of bloom filter~\cite{bloom_1970_bloom_filters} made using a fixed-length truncated hash: two identical values always return the same blind index value; however two different values may also give an identical blind index value.
By using blind indices of sufficient length, we can meaningfully filter data in a large database table.
By keeping this length low enough, we can maintain a sufficiently high number of \emph{collisions} (false positives), that prevents an adversary from inferring equality of values.

\sys needs to decrypt and encrypt data, and it must be protected against its own environment (\ie no adversary must be able to extract keys from its memory).
To enforce such guarantees, we leverage SEV-SNP~\cite{kaplan_amd_nodate}, a trusted execution environment (TEE) offered on modern AMD EPYC server-grade CPUs, and widely available for use in cloud providers.

TEEs are hardware-protected memory areas (often referred to as \emph{enclaves}) that are fully isolated from the host operating system.
SEV-SNP is a virtual-machine based TEE, which means it runs VMs with encrypted memory, protected execution state (CPU registers), and strong integrity protection.
Hence, malicious hosts/hypervisors cannot read nor write in the memory of a confidential SEV-SNP VM.
In addition, TEEs offer multiple ways for external observers to \emph{attest} that a particular piece of software is indeed running in a TEE (and not in an untrusted environment), and that it has not been altered in any way.

\textbf{Roadmap.}
In \S\ref{sec:related}, we survey related work on protected database systems.
In \S\ref{sec:terminology}, we introduce the terminology of the different components that intervene in a typical modern internet application.
\S\ref{sec:approach} presents the architecture of \sys.
Our security analysis of \sys is presented in \S\ref{sec:security-analysis}.
We present the experimental evaluation of \sys in \S\ref{sec:results}, before concluding in \S\ref{sec:future}.

\begin{table*}[!t]
    \scriptsize
    \centering
    \setlength{\tabcolsep}{0.5pt}
    \setlength{\extrarowheight}{3pt}  \rowcolors{1}{gray!10}{gray!0}
    \begin{tabularx}{\textwidth}{X|*{6}{q{13mm}}|q{9mm}}
        \toprule
        \rowcolor{gray!25}
        & \makecell[c]{CryptDB\\\cite{popa_cryptdb_2011}} & \makecell[c]{Crypt-\\SQLite\\\cite{DBLP:conf/nana/WangLSMWYSLZD17}} & \makecell[c]{Enclave\\DB\\\cite{priebe_enclavedb_2018}} & \makecell[c]{Always\\Encrypted\\\cite{DBLP:conf/sigmod/AntonopoulosASE20}} & \makecell[c]{Gabel\\\textit{et al.}\\\cite{DBLP:conf/cbms/GabelM17}} & \makecell[c]{StealthDB\\\cite{DBLP:journals/popets/VinayagamurthyG19}} & \makecell[c]{This\\paper}\\
        \midrule
        Can run TPC-C                   & \compfull & \compfull & \compfull & \compfull & \compnone & \compfull & \compnone\\ Query public/private data       & \compfull & \compnone & \compnone & \compfull & \compnone & \compfull & \compfull\\ Per-user/app keys& \compfull    & \compfull & \compnone & \compfull & \compnone & \compnone & \compfull\\
Unmodified DBMS& \compfull      & \compnone & \compnone & \compnone & \compfull & \compnone & \compfull\\
        DBMS outside TCB& \compfull     & \compnone & \compnone & \compfull & \compfull & \compfull & \compfull\\
        Avoid OPE       & \compnone & --- & --- & \compnone & \compfull & \compfull & \compfull\\
        Avoid deterministic enc.        & \compnone & --- & --- & \compnone & \compnone & \compfull& \compfull\\
        End-to-end encryption           & \compnone & \compnone & \compnone & \compnone & \compnone & \comppart & \compfull\\ Encryption granularity          & column & database & table & column & table & column & column\\
        Supported TEEs                  & --- & SGX & SGX & SGX & SGX & SGX & SEV\\
        \bottomrule
    \end{tabularx}
    \vspace*{2mm}
    \caption{\label{table:rw}Comparison of the state-of-the-art protected databases.}
    \vspace*{-20pt}
    \scriptsize

\end{table*}

\section{Related Work}
\label{sec:related}

We survey state-of-the-art protected database solutions, comparing their cornerstone features in \Cref{table:rw}.
Each feature is assessed as either non-applicable (---), missing (\compnone), partially (\comppart), or fully (\compfull) available.
We consider their support for TPC-C, if queries can combine non-encrypted public data and encrypted sensitive data, if the DBMS supports per-user app keys or if it required modifications, and if they use any TEE.
We include potential native support for order-preserving encryption (OPE)~\cite{grubbs_learning_2019}, or if the DBMS avoids deterministic encryption due to known security issues~\cite{grubbs_learning_2019}.
We distinguish between systems with end-to-end encryption schemes for data stored in the database.
Partial availability (\raisebox{-0.5ex}{\comppart}) signifies that the application backend must be trusted, and full availability (\raisebox{-0.5ex}{\compfull}) indicates that only the end-user device requires trust.

We observe the following.
Database systems can be protected by software and hardware-based techniques~\cite{fuller_sok_2017}, extensively explored by academia and industry.
CryptDB~\cite{popa_cryptdb_2011} uses a trusted proxy between clients and the database system.
This approach offers the benefit of abstraction, allowing the proxy to interface with various database engines seamlessly.
\sys follows a similar approach for its trusted proxy.
Hardware-assisted TEEs offer strong security guarantees, protecting data confidentiality and integrity even when hosted in untrusted environments, tackling a more powerful threat model than CryptDB.
However, we observe how most of them lack abstraction capabilities, requiring tight coupling with specific database engines, or failed to provide robust end-to-end encryption.
Authors in~\cite{grubbs_learning_2019} showed vulnerabilities of CryptDB's encryption schemes, such as order-preserving and deterministic encryption, to approximate database recovery attacks.
Order-preserving encryption maintains the plaintext order in the ciphertext, and deterministic encryption produces the same ciphertext for identical plaintexts when using the same encryption scheme repeatedly.
To address these security concerns, \sys combines the benefits of a proxy-based architecture with a stronger threat model.
Our approach ensures end-to-end encryption to protect user data and leverages TEEs to provide confidentiality, integrity and trust in stored data.

The execution of database systems within TEEs is challenging, due to the inherent constraints of these secure environments.
Consequently, two main approaches have emerged to design such systems.
The first approach involves fully encapsulating the database system within the TEE, exposing its services through secure communication channels (\ie network interfaces).
Alternatively, one can partition the DBMS, shielding only critical components within the trusted environment.
The choice between these two implementation strategies is a subject of ongoing debate~\cite{DBLP:conf/usenix/LindPMOAKRGEKFP17,DBLP:conf/middleware/YuhalaMFST0GL21}, as it represents a fundamental tradeoff between minimizing the trusted computing base (TCB) to reduce the attack surface, and the ease of deploying off-the-shelf database systems with modified interfaces.
While a smaller TCB enhances security by limiting potential vulnerabilities, the latter approach simplifies adopting existing database solutions in TEE-protected architectures.

CryptSQLite~\cite{DBLP:conf/nana/WangLSMWYSLZD17} ensures data confidentiality and integrity by fully encapsulating the database system within SGX enclaves using AES-GCM 128-bit encryption for each database page.
In contrast, \sys minimizes encryption operations by supporting protected columns, thereby reducing the performance impact.

EnclaveDB~\cite{priebe_enclavedb_2018} can manage both public and sensitive data, storing the latter within an SGX enclave using table-level encryption granularity.
It leverages a modified version of Hekaton~\cite{DBLP:conf/sigmod/DiaconuFILMSVZ13} for secure data management within the TEE and establishes secure communication channels.
However, EnclaveDB requires a trusted client machine to compile database queries, aiming to minimize the TCB at the cost of client-side modifications.
Our system, on the other hand, supports column-level encryption granularity for sensitive data and enables the processing of both public and sensitive data within a single query, addressing a limitation of EnclaveDB.
Furthermore, \sys parses standard SQL queries within the enclave without requiring a database engine inside the TEE, further reducing the TCB size.

Always Encrypted (AE)~\cite{DBLP:conf/sigmod/AntonopoulosASE20} extends Microsoft SQL Server to store encrypted data with column-level granularity in the regular database engine.
It notably uses SGX enclaves to execute queries on encrypted data, decrypting it only within the enclave memory.
\sys introduces a proxy that abstracts the underlying database engine, enabling adaptability to various database systems.
Moreover, our solution realizes an end-to-end encryption scheme, ensuring that data is never decrypted on the same infrastructure hosting the database engine.

Gabel and Mechler's secure database outsourcing approach~\cite{DBLP:conf/cbms/GabelM17} shares similarities with \sys, using an SGX enclave to host a proxy that intercepts client-database communication.
They protect sensitive data tables by concatenating and encrypting each row's values, storing the resulting ciphertext in a single column while leaving the row identifier unencrypted.
In contrast, our work encrypts columns individually, eliminating the need to decrypt entire rows when accessing a subset of columns, thereby improving querying efficiency.
Our end-to-end encryption scheme uses per-user keys, ensuring secure communication between clients and the proxy, while~\cite{DBLP:conf/cbms/GabelM17} does not mention this security aspect when secured within SGX enclaves.
In addition, our encryption scheme selectively encrypts sensitive data, avoiding the encryption of the entire dataset, as instead required by~\cite{DBLP:conf/cbms/GabelM17}.

StealthDB~\cite{DBLP:journals/popets/VinayagamurthyG19} relies on proxies inside an Intel SGX enclave, separating responsibilities among multiple enclaves for authentication, query preprocessing, and database operations.
It supports column-level encryption granularity and introduces new encrypted data types and functions as extensions to the underlying PostgreSQL database engine.
\sys supports a stronger threat model, by introducing blind indexes that obfuscate access patterns and prevent leakage of sensitive data from the index structure, such as record ordering. Our solution offers better end-to-end security guarantees than StealthDB: client data is encrypted locally (\eg from within a web browser) via  per-user keys, similar to privacy-focused products \cite{proton_e2ee, bitwarden_e2ee}.
Our approach is DBMS-agnostic and it avoids the need for new data types, unlike StealthDB's engine-specific extensions to PostgreSQL.

\section{System Model}\label{sec:terminology}

We consider the following three roles: the \emph{end-user}, the \emph{service provider} and the \emph{database administrator}.
The end-user is the data owner and client of the system, intending to upload data in the system.
The service-provider builds, distributes, and sells access to the application.
Finally, the database administrator hosts and manages the database. These roles can be shared, \eg a single entity can act simultaneously as service provider and database provider.

We model a typical internet application using the following three components: the \emph{client}, the \emph{application backend}, and the \emph{database}.
The client is the interface used by the end-user to access the application, \ie a website or software to install on their device.
The client communicates with the application backend operated by the service-provider, handling the core business logic.
In turn, the application backend stores its data in a database management system (DBMS), hosted and maintained by the database administrator.
Note that while we build \sys atop the MySQL DMBS, the architecture is flexible and can be easily ported to alternative SQL systems.

We introduce a fourth component, the \emph{database proxy}, sitting between the application backend and the database.
It handles transparent encryption and decryption of data and may rewrite queries.
This component is detailed in \S\ref{sec:approach}.

\sys provides \emph{end-to-end encryption}, \ie some data can only be decrypted by the client and database proxy.
The database proxy is a trusted application running in a TEE, and as long as it does not reveal user data, the data is effectively only accessible by the \emph{client}.
We provide a detailed security analysis in \S\ref{sec:security-analysis}.

\section{The \sys Database Proxy}
\label{sec:approach}

\subsubsection{Design goals.}
\label{subsec:design-goals}
Our database encryption system has the following main design goals.

\paragraph{(a) End-to-end data security:} The end user's confidential data should only be accessible to the user and the proxy.
\paragraph{(b) Ease of implementation:} The approach should require minimal modification to the client and application backend. This is partially achieved by distributing client libraries for the proxy system. \paragraph{(c) Efficient use of the DBMS capabilities:} Wherever possible, \sys should rely on the DBMS native capabilities for filtering data.

\begin{figure}[!t]
    \centering
    \includegraphics[scale=0.8]{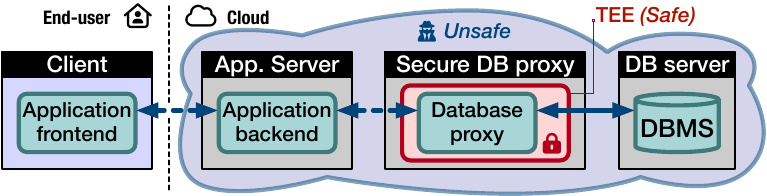}
    \caption{Architecture of \sys. Dashed arrows denote data encrypted using a session key, the full arrow denotes data encrypted using a long-term key.}
    \label{fig:components}
    \vspace*{-10pt}
\end{figure}
\subsubsection{Overview.}
\label{subsec:overview}
Figure~\ref{fig:components} shows the architecture of \sys, including the flows of data across the components.
The \emph{end-user} interacts with the \emph{client}, which encrypts and sends queries to the \emph{application backend}, receives and decrypts their results, and present them to the \emph{end-user}.
The queries are encoded using an application specific serialization format (\eg, JSON or XML), and transmitted using an application specific protocol (\ie, HTTP).
Only part of a query or response is typically sensitive: the \emph{application backend} may need a cleartext view of some parts of a query or response to correctly operate (\eg, to check for correct permissions, to send emails).
Hence, only some fields are encrypted in the queries and responses, matching the encrypted columns in the database.
The schema of the query is transmitted in clear.

The \emph{application backend} receives partially encrypted queries from the \emph{client}, applies business logic based on the cleartext fields, then transmits SQL queries to the \emph{proxy}. The SQL queries may contain encrypted values extracted from the client query, and the results to the queries contain a mix of encrypted and cleartext values.
The backend can also apply business logic based on the cleartext parts of the SQL response, before retransmitting data back to the client.
As with the request, encrypted and cleartext values are mixed in the application specific data serialization format.
Only minimal modifications are required in the backend (\eg for login and registration), achieving design goal (b).

The \emph{proxy} receives partially encrypted SQL queries from the \emph{application backend}, communicates with the \emph{database server}, and generates partially encrypted SQL responses.
When it receives a simple non encrypted query, the proxy trivially forwards it to the database server, and directly forwards the response to the client, without any further processing.
If the query contains encrypted fields, tries to access encrypted data, or uses one of the custom functions of the proxy, then the proxy must handle it (see \S\ref{subsec:query-processing}). The proxy may encrypt or decrypt data, and may submit additional SQL queries to the \emph{database server}.
The use of encryption achieves our design goal (a), and the use of blind indices design goal (c).
The rationale for encryption is further detailed in \S\ref{subsec:security-measures}.

Note that all communication between the two trusted components (the \emph{client} and the \emph{application proxy}) goes through the \emph{application backend}.
Moreover, all encryption happens at the level of individual values, while the communication channel is not protected.
For a single client request containing a single encrypted value, the \emph{application backend} may issue multiple SQL queries containing that same value.

\subsubsection{Establishing a trusted secure channel.}
\label{subsec:establishing-a-trusted-channel}
\begin{figure}[!t]
    \centering
    \includegraphics[scale=0.8]{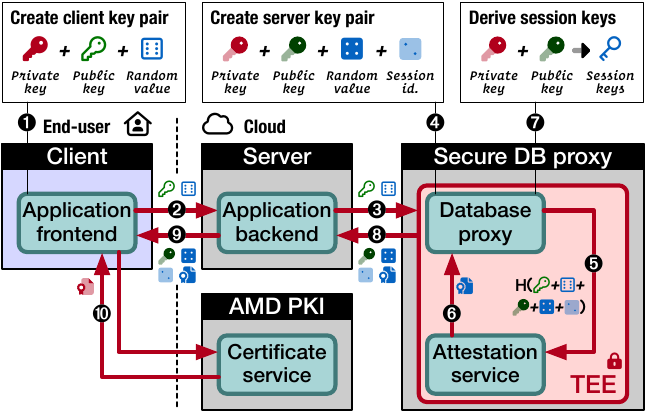}
    \caption{Exchange of packets during session establishment}
    \label{fig:session_establishment}
    \vspace*{-10pt}
\end{figure}

Upon start, the \emph{client} ensures it can trust the \emph{proxy} and establishes an encrypted tunnel with it.
These two steps are implemented as a single key exchange, similar to TLS~\cite{rescorla_transport_2018}.
Since the \emph{application backend} only interacts with the \emph{proxy} via SQL, the key exchange is implemented as a custom SQL function, \texttt{KEY\_EXCHANGE}, which is intercepted by the proxy.
Figure~\ref{fig:session_establishment} illustrates the protocol.

To initiate the key exchange, the \emph{client} generates a random value and an ephemeral keypair (\ding{202}).
It encodes the public key and random value and sends them to the \emph{application backend} (\ding{203}), which sends them to the \emph{proxy} using the custom SQL query (\ding{204}).
The proxy generates its own random value, ephemeral keypair, and a session ID to identify this particular session in further communication (\ding{205}).
To attest its trustworthiness, the proxy then requests an attestation to the TEE secure processor, passing a hash of both random values, both public keys, and the session ID, as custom data (\ding{206}--\ding{207}).

To complete the key exchange, the \emph{client} and \emph{proxy} each combine their own ephemeral private key with the other party's ephemeral public key, and derive two final \emph{session keys}, one for each protocol direction (client to proxy and proxy to client), which they store in memory (\ding{208}).
Finally the response is sent back to the backend (\ding{209}) and the frontend (\ding{210}).
In addition, the \emph{client} needs to assess the trustworthiness of the \emph{proxy}.
To do so, it fetches the root certificates and hardware public key from the TEE vendor (\ding{211}), and ensures the attestation returned by the \emph{proxy} matches the key exchange, to be signed by the correct public key.

Because the client typically runs in a browser, we choose cryptographic primitives that are available in the Web Cryptography API~\cite{huigens_web_nodate}: we use Elliptic-Curve Diffie-Hellman (ECDH) on the NIST P-256 curve~\cite{technology_digital_2023} for the key exchange, we derive the key using HKDF~\cite{krawczyk_cryptographic_2010} instantiated with SHA-256. Finally we encrypt subsequent messages exchanged in the session with AES-GCM.

\subsubsection{Obtaining long-term keys.}
\label{subsec:key-management}
The \emph{session keys} used to encrypt data in transit between them for the duration of a communication session have a short lifetime (\ie a few hours).
For long-term storage in the database, the proxy derives a different set of \emph{long-term} keys, that are preserved across sessions.
For each user, the proxy stores in the database an encrypted master long-term key, which can only be unlocked by logging in on an established session.
Then, each confidential column in the database is encrypted using a different key, derived from such master long-term key.
After successfully initiating a session, the client can use the custom procedures \texttt{REGISTER} and \texttt{LOGIN}, to register a long-term key and unlock a previously stored long-term key, respectively.
Both procedures take as arguments a username and a password, the latter encrypted using the session key.

\paragraph{Registration.} To register a user given its username and password, the proxy generates the master long-term key randomly, then encrypts it using a key derived from the user password and a salt, and finally stores the encrypted key and salt in the database.
This method simplifies user password's changes, as only the master key needs to be re-encrypted.
To derive the temporary key from the password, we use the Argon2~\cite{biryukov_argon2_2017} PBKDF, with Chacha20-Poly1305~\cite{bernstein_chacha_2008,bernstein_poly1305-aes_2005} AEAD cipher to encrypt the long-term key.

\paragraph{Login.} To log a user in, the proxy retrieves its associated record (salt and encrypted master key) from the database, recomputes the intermediate key using the password and salt, and uses that to decrypt the long-term key.
On a successful login, the proxy associates the decrypted long-term key with the session ID and keeps this mapping in secure memory during the whole session.
It also returns a success response code to the application backend, used to log the user in the application with the same request.

\subsubsection{Persistent storage and blind indices.}
\label{subsec:persistent-storage}
\sys uses two different sets of keys: session keys to encrypt data in transit between the client and proxy, and long-term keys to encrypt data at rest in the database.
We detail here how we encrypt the confidential data and how to derive blind indices, thus enabling processing some filter queries directly in the database without leaking sensitive data.

\paragraph{Encrypting confidential columns.} Data stored in a confidential column is encrypted using a key specific to the user and the column in which the data is inserted.
That key is derived using HKDF~\cite{krawczyk_cryptographic_2010} with SHA-256, using the user long-term key as the key, and the table and column names as \emph{info}.
The chosen encryption scheme guarantees \emph{semantic security}, meaning that multiple encryptions of the same plaintext with the same key give different ciphertexts.
In addition, it is authenticated, so any modification to the ciphertext prevents future decryption.
The single-use nonce is randomly generated when encrypting data and added as prefix to the encrypted value.

\paragraph{Blind indices.} Because the encryption scheme preserves semantic security, it is not feasible to use the encrypted columns directly to filter data, even for simple equality queries.
A simple solution to this problem is to process filters over encrypted data entirely in the proxy, by decrypting all rows and only returning those for which the plaintext matches the filters.
This approach works well when the number of records to filter is small, \eg because the plaintext filters in the query reduced that number.
However, this solution does not fully leverage the capabilities of the database server to efficiently filter data.

A more efficient solution is to insert additional data alongside encrypted columns, to allow the database server to filter the data partially, called \emph{blind indices}\cite{noauthor_ciphersweet_nodate}.
A blind index is a hash derived from the user key and the plaintext value and is truncated, hence multiple different plaintexts can give the same blind-index value.
Given two identical blind-index values, an adversary cannot confirm if the original plaintexts are also identical.
When data is inserted or updated in the database, the associated blind-indices are automatically computed by \sys and transparently added to the query.
When querying over an encrypted column, \sys computes the associated blind index and replaces it in the query.
It then decrypts the values and re-filters the plaintexts, as multiple plaintexts give the same blind index.

The size of the blind index is determined manually on a per-column basis, as a function of the size of the plaintext space of that column.
Given a blind index of size $n$ bits with $r$ records, the average number of collisions (that is, identical blind index values for different plaintexts) is $r \times 2^{-n}$, assuming that the size of the input space is itself bigger than $2^n$.
To achieve an average number of collision $C \geq 2$ over $r$ records or more, we should therefore have $n~\geq~\log_2r~-~\log_2C$ bits.
The average number of collisions should be kept higher than 2 for security, and lower than $\sqrt{r}$ to meaningfully impact filtering performance~\cite{noauthor_ciphersweet_nodate}.

\subsubsection{Query processing.}
\label{subsec:query-processing}
When the proxy receives a query, it parses it and accesses its configuration to check if it operates over encrypted columns.
If so, the proxy identifies the client from which the query originates, decrypts and re-encrypts its data, sends it to the database, and then decrypts and re-encrypts its results.

Internally, the response parsing is implemented using an iterator.
Whenever the client is ready to receive a new row, it is pulled from the iterator, which reads the next row from the database, decrypts it, applies filters, and then re-encrypts it.
Using iterators allows to process arbitrary numbers of rows without being bounded by the proxy's memory.
It also reduces latency, as the proxy can send the first row without waiting on all subsequent ones.

\paragraph{Retrieving the long-term keys from a query.}
A single proxy is designed to handle thousands of end-users connected at the same time.
It must discern between the different end-users issuing queries, to encrypt and decrypt values for its owner and to recover the correct long-term key.
A 64 bits \emph{session ID} is issued to the client during key-establishment, and associated with the long-term key at login.
That session ID is additionally used as associated data when authenticating each encrypted value in each query.

To associate a request with a session ID, we offer two options: prefixing the encrypted values, or providing the ID directly.
In the first case, the \emph{client}, when it performs a query with confidential data, prefixes each encrypted value with its session ID before transmitting them.
The proxy then tries to recover a session ID in each encrypted value it receives.
In the second case, the \emph{application backend} appends a special function call to its query, \texttt{SESSION\_ID(\textit{sid})}.
The proxy detects this function in the \texttt{WHERE} clause of a query and removes it.

We observe that the proxy should only decrypt values for confidential columns.
If the proxy decrypts all received encrypted values, it may write a decrypted value to a cleartext column.
Similarly, the proxy should only allow filtering over encrypted columns using encrypted values, to prevent an adversary from verifying if a value is present in the encrypted values and to which record it corresponds.

\paragraph{Double-filtering.}
When the proxy receives a \texttt{SELECT} or \texttt{UPDATE} query that filters over encrypted columns, it needs to filter the data in two steps.
First, the filters need to be replaced with their blind indices equivalents (if available) in the request transmitted to the database.
Second, the rows returned by the database need to be re-filtered by the proxy.
To ensure the practicality of the second step, the proxy transparently adds the columns used in the \texttt{WHERE} clause of the query to its projection, ensuring these columns will be returned by the database server.

We note that this filtering method prevents computing aggregation operations (\eg \texttt{SUM}, \texttt{COUNT}, \texttt{AVG}) directly in the database server if the filters include encrypted columns, as the database server will include values that should be excluded in its aggregate.
For \texttt{UPDATE} queries, we transform them into a transaction: first select the individual identifiers of all the rows matching the filter, and then update rows matching these identifiers.

\section{Security Analysis}
\label{sec:security-analysis}

\subsubsection{Threat model.}
\label{subsec:threat-model}
We assume a powerful adversary with entire control over the software and hardware stack.
His goal is to gain information about the confidential data transmitted by the \emph{end user}.
Denial of service attacks, or other attacks altering the correct behavior of the application, such as dropping queries, cloning encrypted data or rolling back data, are out of scope.
These attacks are generally addressed by orthogonal solutions, \eg monotonic counters~\cite{DBLP:conf/dsn/GregorOVPQAMSFF20}.
In the following, we further refine the threat model.

\emph{DBMS.} The DBMS and the server on which it runs are entirely untrusted.
An adversarial database administrator can execute arbitrary read and write queries on stored data, and may also use the application as an end-user to try to access data.

\emph{Database proxy.} The database proxy runs in a TEE.
The adversary has access to the physical machine on which the TEE runs, and can modify the disk image of the virtual machine before launching it.
The TEE prevents the adversary from reading or modifying the memory of the virtual machine while it's running.
Side-channel attacks targeting TEEs are outside the scope of our threat model, and mitigations exist~\cite{DBLP:conf/uss/LiZWLC21,DBLP:conf/sp/LiWW0TZ22,DBLP:conf/dimva/WangLZL23,Zhang24,Schluter24Heckler,Schluter24WeSee}.

\emph{Application backend.} The application backend is fully untrusted.
An adversarial application developer can inspect any data going through the application backend, and can implement arbitrary modifications to the backend.

\emph{Client.}
The client code can be audited by the end user before execution and is therefore entirely trusted.
Attacks targeting the client code are out of scope of this threat-model.

\subsubsection{Security measures.}
\label{subsec:security-measures}
Considering the aforementioned thread model, our approach implements the following security measures.

\emph{Data at rest.}
Confidentiality of data stored in the DBMS is ensured by the use of a semantically secure authenticated encryption scheme, ChaCha20-Poly1305.
The security of the encryption keys depends on the strength of the user's password, but the use of a memory-hard key derivation function, Argon2, makes offline brute force attacks inefficient.
The use of incorrectly sized blind indices may reveal when two plaintexts are equal, therefore boundaries specified in \S\ref{subsec:persistent-storage} must be respected to reduce this risk.
The use of different keys for different users prevents confused deputy attacks in which the adversary changes the user associated with a record to gain access to it.

\emph{Data in the proxy.}
The proxy is a critical component as it accesses the session keys and long-term keys of any logged-in user.
It is protected from its environment by the use of the AMD SEV-SNP~\cite{kaplan_amd_nodate} TEE, which prevents compromised OS processes from accessing its memory and, therefore, the keys or confidential data.
The risk of accidentally leaking cryptographic materials through programming errors is reduced by using a memory-safe language (Rust) and by relying on standard and audited cryptography libraries.
In addition, the proxy only keeps keys in memory for users currently logged in, reducing the impact of a critical TEE vulnerability to only those users.

\emph{Key exchange.}
Its role is both to establish a secure channel between the client and proxy, and for the client to establish trust in the proxy before exchanging sensitive data.
We establish trust in the content of the virtual hard drive, including the OS and system files, and the compiled code of the database proxy.
The integrity of these components is asserted through remote attestation~\cite{DBLP:conf/dais/MenetreyGKPFSR22}, and the client contains the necessary code measurements to assess the authenticity of the proxy. A hash of the entire key exchange is included in the attestation to prevent man in the middle attacks.

\emph{Data in transit.}
Data in transit between the client and the proxy is encrypted using the AES cipher in GCM mode.
To prevent catastrophic nonce reuse, each direction of transit uses a different key and a counter which is incremented for each encrypted value.
The client does not exchange any confidential data with the proxy until it has completed the key exchange and appraised the attestation.

\section{Evaluation}
\label{sec:results}
We present here our extensive experimental evaluation of \sys.
We implemented our prototype in 5500 lines of Rust, used for its low level interface and memory safety features, both desirable in a TEE.
The proxy uses generic data types which are DBMS agnostic, and a DBMS specific implementation layer that translates these data-types to the underlying DBMS protocol.
We demonstrate an integration with the MySQL protocol running in the AMD SEV-SNP~\cite{kaplan_amd_nodate} TEE for its guest attestation features. Approximately 1700 lines of code are specific to the MySQL implementation, and about 50 are specific to AMD SEV, including the attestation data structure.

\subsubsection{Experimental setup.}
Two servers were used: an AMD EPYC 9124 (16 cores, 3\,GHz) hosting the database and \sys, and an AMD EPYC 7302P (16 cores, 3\,GHz) hosting the application backend and benchmarking client. MySQL 8.2 ran in a Docker container; \sys operated in an 8-core AMD SEV-SNP VM with Ubuntu 24.04 LTS and AMD-SEV Linux kernel 6.8.0-35. NodeJS 20.15 was used for the application backend and benchmarking suite.

\subsubsection{End-to-end performance tests.}
\label{subsec:end-to-end-benchmarks}
We simulated a real-world use case with the components in \S\ref{sec:terminology} using a custom-built NodeJS server application and benchmarking client for managing patients.
We generated test data including confidential names and social security numbers (SSN), encrypted in the tests.

\begin{figure}[!t]
    \includegraphics[width=\textwidth]{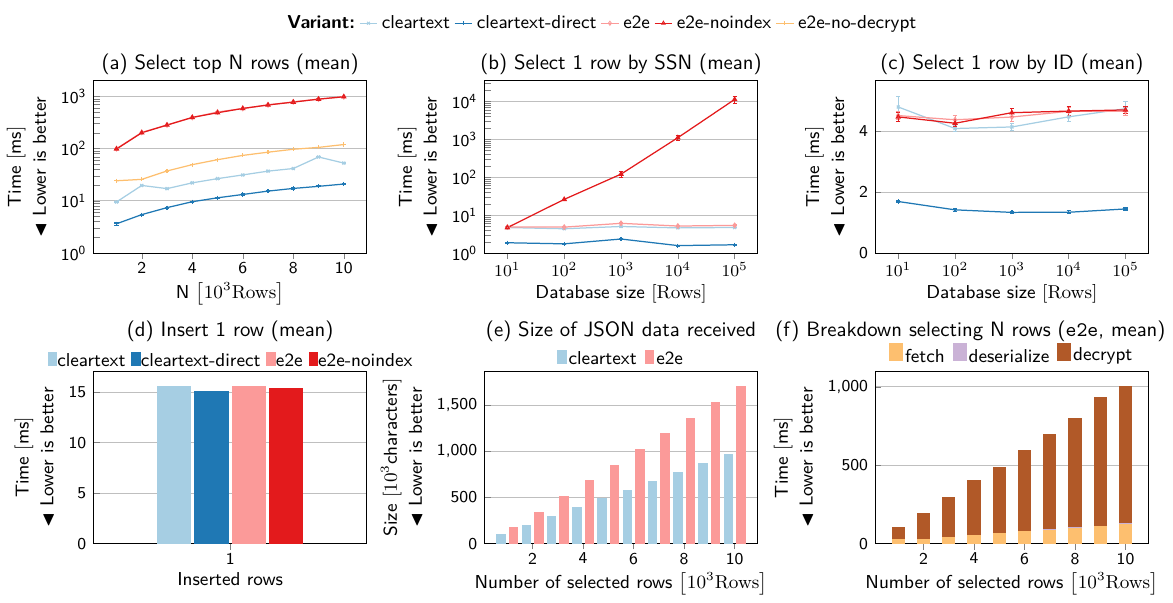}
    \caption{Results of end-to-end performance test}
    \label{fig:e2e_benchmarks}
    \vspace*{-10pt}
\end{figure}

Figure~\ref{fig:e2e_benchmarks} (a)-(e) compares the cleartext app (\texttt{cleartext-direct}), its variant through \sys without encryption (\texttt{cleartext}), and our encryption proxy with (\texttt{e2e}) and without (\texttt{e2e-noindex}) SSN blind indexing.
Graph (a) includes \texttt{e2e-no-decrypt}, a variant of \texttt{e2e} where the client does not decrypt the data after receiving it, which isolates proxy performance.
We measure the time between request encryption and issuance, and between response reception and decryption.

Graph (a) shows a \texttt{SELECT * \dots\ LIMIT N} query with no filter and a random offset.
Total time grows linearly with the number of rows returned for all variants.
Client-side decryption dominates \texttt{e2e} time, confirmed by graph (f), which breaks down the time for selecting $N$ rows with \texttt{e2e} into HTTP query \texttt{fetch}, JSON deserialization \texttt{deserialize}, and client-side decryption \texttt{decrypt}.
The \texttt{e2e-no-decrypt} variant, which removes client decryption time, shows an average overhead of 115\% compared to \texttt{cleartext}, or 7 \si{\micro\second} per row. This overhead is likely dominated by row parsing and encryption, over initial query parsing.
Comparing \texttt{cleartext} to \texttt{cleartext-direct} shows an average overhead of 169\% (4 \si{\micro\second} per row) for the proxy processing and additional network roundtrips.
The complete overhead of the encrypted proxy (\texttt{e2e-no-decrypt} compared to \texttt{cleartext-direct}) is therefore 462\%, or 11 \si{\micro\second} per row.
Running \sys outside of SEV, we observed a much smaller overhead of 54\% between \texttt{cleartext} and \texttt{cleartext-direct}, suggesting that the cost of emulation and of SEV in particular is a big part of the overall overhead of our system.

Plots (b) and (c) represent \texttt{SELECT *} queries with filters on SSN and ID, respectively, returning a single row.
Except for \texttt{e2e-noindex}, query time remains constant in both tests, independent of database size.
Comparing \texttt{e2e} and \texttt{cleartext} in (c) shows a 3\% average slowdown (0.12 \si{\milli\second}), while comparing with direct database access \texttt{cleartext-direct} shows a 237\% slowdown (3.19 \si{\milli\second}), suggesting query processing by \sys dominates the overhead.
Graph (b) demonstrates the usefulness of blind indices when filtering over an encrypted column: in \texttt{e2e-noindex}, \sys retrieves and decrypts all rows to filter them, and the variant's runtime growing linearly with database size.
Meanwhile in \texttt{e2e}, only rows that match the blind index are decrypted by the proxy, and the runtime remains constant.

Plot (d) shows an \texttt{INSERT} query for a single row.
All variants achieve comparable times within the margin of error, suggesting that the base cost of inserting a row in the SQL database dominates.

Finally, in (e), we present the JSON payload size received by the client for cleartext or encrypted queries.
The response comprises a JSON array with fields \texttt{id, doctorOfficeId, name, SSN}.
Encryption consistently increases body size by 75\%.
This overhead is expected to decrease for larger encrypted values due to the fixed-size prefix nonce.
However, base64 encoding inherently imposes a minimum 33\% size increase.

\begin{figure}[!t]
    \includegraphics[width=\textwidth]{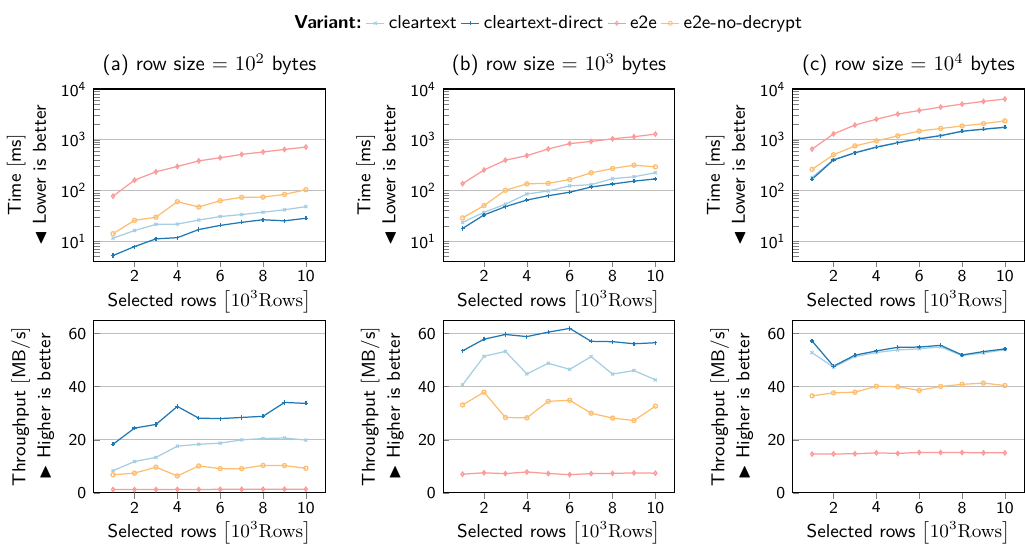}
    \caption{End-to-end test results in various plaintext data sizes (expressed in bytes)}
    \label{fig:macro_datasize_grid}
    \vspace*{-10pt}
\end{figure}

The previous tests use small encrypted columns, each $<$\,20 characters, and overhead is dominated by fixed costs such as query parsing and general row processing.
\Cref{fig:macro_datasize_grid} illustrates the results of selecting $N$ rows of varying sizes using the previously described variants to show how the overhead evolves with data size.

\texttt{e2e-no-decrypt}'s average overhead compared to \texttt{cleartext-direct} decreases from 225.4\% for datasize $10^2$ to just 36.1\% for datasize $10^4$.
The overhead of \texttt{cleartext} compared to \texttt{cleartext-direct}, that is the pure overhead of our system without any encryption, also decreases with data size, from +60\% for size $10^2$ to 2\% for size $10^4$.
This indicates that higher data-sizes reduce the impact of \sys's query processing and additional round trips, and the remaining encrypted overhead can be primarily attributed to decryption and encryption of data.

\subsubsection{Micro-benchmarks.}
\label{subsec:precise-benchmarks}
\begin{figure}[!t]
    \includegraphics[width=\textwidth]{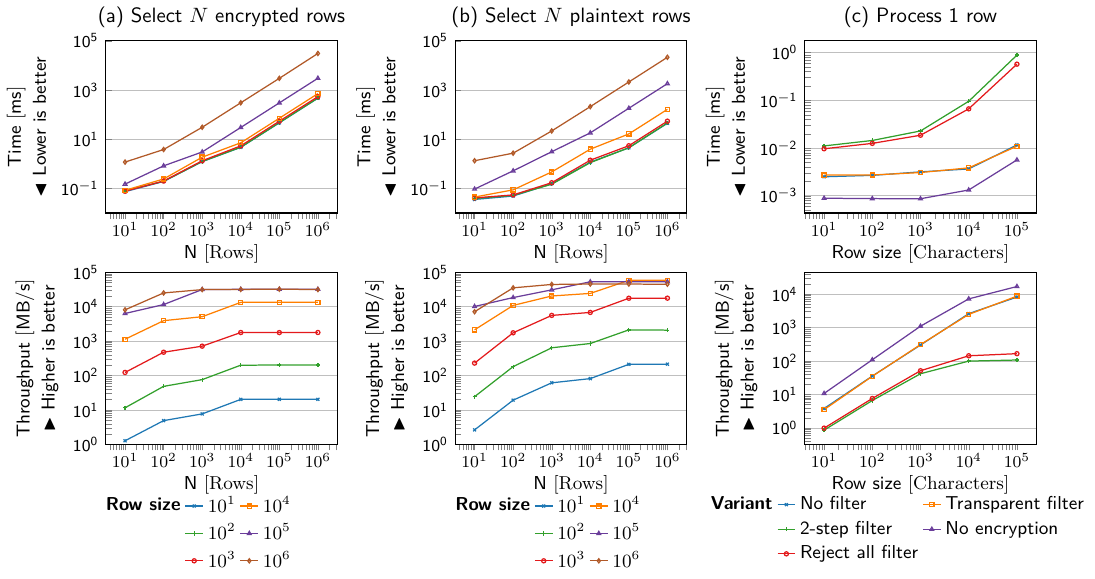}
    \caption{Results of our micro-benchmarks}
    \label{fig:microbenchmarks}
    \vspace*{-10pt}
\end{figure}

To better understand how \sys behaves in certain specific conditions, we carried micro-benchmarks on the code handling the selection of rows.
We present our results in \Cref{fig:microbenchmarks}.

In (a) and (b), we present the time and throughput for a \texttt{SELECT} query that returns $N$ rows of different \textit{row size}, respectively encrypted and in cleartext.
To isolate the proxy overhead, we replace the underlying database call with an injected crafted response containing randomized data of the correct size.
Because we test in isolation, we do not stream the results to the client, but instead iterate on each of them to consume the stream.

In (c), we test how long it takes our filtering and encryption/decryption iterator to process a single row, after its packet has been parsed but before it is passed to the caller, in different configurations.
In the \texttt{No filter} variant, rows are decrypted and re-encrypted, but no filtering or projection is performed.
The \texttt{Transparent filter} variant simulates a case of a query with a cleartext filter, so rows are decrypted, passed to a filter that does nothing, then re-encrypted.
The \texttt{2-step Filter} variant presents a query with an encrypted filter, in which rows are decrypted, then compared to a decrypted value, then re-encrypted.
The counterpart is the \texttt{Reject all filter} variant, which is similar but in which the comparison value is different from the compared value, and so all rows are rejected.
Finally, the \texttt{No encryption} variant quantifies the overhead of our custom iterator when no encryption or decryption operation happens.

We observe in both (a) and (b) that the processing time of the proxy mostly depends on the number of rows.
This matches expectations and end-to-end tests, as each protocol packet must be parsed and its content decrypted and re-encrypted (in graph (a)), which leads to a per-row overhead.
In addition, the throughput graphs reveal that this per-row cost is the dominant cost, as multiplying the row-size by ten roughly multiplies the throughput by the same factor, although this seems to reach a ceiling at $10^5$ sized rows.

Graph (c) reveals the breakdown of the different costs involved in encryption and two-step filtering of a single row.
As expected, using no encryption nor filtering gives the highest throughput of all variants.
Variants with encryption/decryption but no filtering come just after.
Finally, variants in which each row must be compared to a value for filtering are the slowest, and their throughput doesn't grow as fast as other variants, as using larger strings causes longer comparison times for each string.

\section{Conclusion and Future Work}
\label{sec:future}

In this paper, we have presented an experimental system that enables encrypting data in an application using user specific keys that are not accessible to the operators of the application.
We have shown how a TEE-enabled proxy can decrypt and re-encrypt this data, without leaking confidential information, in order to compute blind indices, which make further retrieval of data more efficient by leveraging the capabilities of the DBMS\@.
We have demonstrated the practicality of \sys and presented some performance data, but because of its limited scope we could not compare our solution with other database protection solutions, thus limiting the relevance of our analysis.

Future work on this topic includes implementing some complex missing aggregations, such as aggregates or joins.
Additionally, we intend to explore the benefits of using a TEE for encryption to implement other features, such as encrypting for groups of users, or computing statistics over encrypted data with differential privacy guarantees.

\subsubsection{\ackname}
This work was supported by the Swiss National Science
Foundation under project P4: Practical Privacy-Preserving
Processing (no. 215216).

\bibliographystyle{splncs04}

\begin{thebibliography}{10}
\providecommand{\url}[1]{\texttt{#1}}
\providecommand{\urlprefix}{URL }
\providecommand{\doi}[1]{https://doi.org/#1}

\bibitem{noauthor_ciphersweet_nodate}
{CipherSweet}: {Searchable} {Encryption} {Doesn}'t {Have} to be {Bitter} -
  {Paragon} {Initiative} {Enterprises} {Blog} (Jan 2019),
  \url{https://paragonie.com/b/HXPUHJZVaub77-Zg}

\bibitem{kaplan_amd_nodate}
{AMD SEV-SNP: Strengthening VM Isolation with Integrity Protection and More}.
  White paper (2020), \url{http://bit.ly/4bJepse}

\bibitem{DBLP:conf/sigmod/AntonopoulosASE20}
Antonopoulos, P., Arasu, A., Singh, K.D., Eguro, K., et~al.: Azure {SQL}
  database always encrypted. In: {SIGMOD}'20 (2020).
  \doi{10.1145/3318464.3386141}

\bibitem{technology_digital_2023}
Barker, E.: Digital signature standard\,{\scriptsize (\MakeUppercase{dss})}.
  NIST  (2013). \doi{10.6028/NIST.FIPS.186-4}

\bibitem{bernstein_poly1305-aes_2005}
Bernstein, D.J.: The {Poly1305}-{AES} {Message}-{Authentication} {Code}. In:
  Fast {Software} {Encryption}. Berlin, Heidelberg (2005).
  \doi{10.1007/11502760_3}

\bibitem{bernstein_chacha_2008}
Bernstein, D.J.: {ChaCha}, a variant of {Salsa20} (2008)

\bibitem{biryukov_argon2_2017}
Biryukov, A., Dinu, D., Khovratovich, D.: Argon2: the memory-hard function for
  password hashing and other applications. Tech. rep. (2017)

\bibitem{bitwarden_e2ee}
{Bitwarden}: Bitwarden security whitepaper. White paper (2021),
  \url{https://bitwarden.com/help/bitwarden-security-white-paper}

\bibitem{bloom_1970_bloom_filters}
Bloom, B.H.: Space/time trade-offs in hash coding with allowable errors.
  Commun. ACM  \textbf{13}(7) (jul 1970)

\bibitem{DBLP:conf/sigmod/DiaconuFILMSVZ13}
Diaconu, C., Freedman, C., Ismert, E., Larson, P., et~al.: {Hekaton}: {SQL}
  server's memory-optimized {OLTP} engine. In: {SIGMOD}'13 (2013).
  \doi{10.1145/2463676.2463710}

\bibitem{fuller_sok_2017}
Fuller, B., Varia, M., Yerukhimovich, A., Shen, E., et~al.: {SoK}:
  {Cryptographically} {Protected} {Database} {Search}. In: {SP}'17 (May 2017).
  \doi{10.1109/SP.2017.10}

\bibitem{DBLP:conf/cbms/GabelM17}
Gabel, M., Mechler, J.: Secure database outsourcing to the cloud:
  Side-channels, counter-measures and trusted execution. In: {CBMS}'17 (2017).
  \doi{10.1109/CBMS.2017.141}

\bibitem{DBLP:conf/dsn/GregorOVPQAMSFF20}
Gregor, F., Ozga, W., Vaucher, S., Pires, R., et~al.: Trust management as a
  service: Enabling trusted execution in the face of byzantine stakeholders.
  In: {DSN}'20 (2020). \doi{10.1109/DSN48063.2020.00063}

\bibitem{grubbs_learning_2019}
Grubbs, P., Lacharite, M.S., Minaud, B., Paterson, K.G.: Learning to
  {Reconstruct}: {Statistical} {Learning} {Theory} and {Encrypted} {Database}
  {Attacks}. In: {SP}'19 (May 2019). \doi{10.1109/SP.2019.00030}

\bibitem{huigens_web_nodate}
Huigens, D.: Web {Cryptography} {API}. {W3C} {Editor}'s {Draft} (Jun 2024),
  \url{https://w3c.github.io/webcrypto/}

\bibitem{krawczyk_cryptographic_2010}
Krawczyk, H.: Cryptographic extraction and key derivation: The {HKDF} scheme
  (2010), \url{https://eprint.iacr.org/2010/264}, cryptology ePrint Archive,
  Paper 2010/264

\bibitem{DBLP:conf/sp/LiWW0TZ22}
Li, M., Wilke, L., Wichelmann, J., Eisenbarth, T., et~al.: A systematic look at
  ciphertext side channels on {AMD} {SEV-SNP}. In: {SP}'22 (2022).
  \doi{10.1109/SP46214.2022.9833768}

\bibitem{DBLP:conf/uss/LiZWLC21}
Li, M., Zhang, Y., Wang, H., Li, K., et~al.: {CIPHERLEAKS:} breaking
  constant-time cryptography on {AMD} {SEV} via the ciphertext side channel.
  In: {USENIX} Security 2021 (2021)

\bibitem{DBLP:conf/usenix/LindPMOAKRGEKFP17}
Lind, J., Priebe, C., Muthukumaran, D., O'Keeffe, D., et~al.: Glamdring:
  Automatic application partitioning for {Intel} {SGX}. In: {ATC}'17. {USENIX}
  (2017). \doi{10.5555/3154690.3154718}

\bibitem{DBLP:conf/dais/MenetreyGKPFSR22}
M{\'{e}}n{\'{e}}trey, J., G{\"{o}}ttel, C., Khurshid, A., Pasin, M., et~al.:
  Attestation mechanisms for trusted execution environments demystified. In:
  {DAIS}'22. Lecture Notes in Computer Science, vol. 13272. Springer (2022).
  \doi{10.1007/978-3-031-16092-9_7}

\bibitem{popa_cryptdb_2011}
Popa, R.A., Redfield, C.M.S., Zeldovich, N., Balakrishnan, H.: {CryptDB}:
  protecting confidentiality with encrypted query processing. In: {SOSP}'11
  (2011). \doi{10.1145/2043556.2043566}

\bibitem{priebe_enclavedb_2018}
Priebe, C., Vaswani, K., Costa, M.: {EnclaveDB}: A secure database using {SGX}.
  In: {SP}'18 (2018). \doi{10.1109/SP.2018.00025}

\bibitem{proton_e2ee}
{Proton}: What is end-to-end encryption and how does it work? (2023),
  \url{https://proton.me/blog/what-is-end-to-end-encryption}

\bibitem{rescorla_transport_2018}
Rescorla, E.: The {Transport} {Layer} {Security} ({TLS}) {Protocol} {Version}
  1.3 (Aug 2018). \doi{10.17487/RFC8446}

\bibitem{Schluter24WeSee}
Schlüter, B., Sridhara, S., Bertschi, A., Shinde, S.: {WeSee}: Using malicious
  \#{VC} interrupts to break {AMD} {SEV-SNP}. In: {SP}'24 (2024).
  \doi{10.1109/SP54263.2024.00262}

\bibitem{Schluter24Heckler}
Schlüter, B., Sridhara, S., Kuhne, M., Bertschi, A., et~al.: {HECKLER}:
  Breaking confidential {VMs} with malicious interrupts. In: {USENIX} Security
  2024 (2024). \doi{10.48550/arXiv.2404.03387}

\bibitem{DBLP:journals/popets/VinayagamurthyG19}
Vinayagamurthy, D., Gribov, A., Gorbunov, S.: {StealthDB}: a scalable encrypted
  database with full {SQL} query support. Proc. Priv. Enhancing Technol.
  \textbf{2019}(3) (2019). \doi{10.2478/POPETS-2019-0052}

\bibitem{DBLP:conf/dimva/WangLZL23}
Wang, W., Li, M., Zhang, Y., Lin, Z.: Pwrleak: Exploiting power reporting
  interface for side-channel attacks on {AMD} {SEV}. In: {DIMVA}'23. Lecture
  Notes in Computer Science, vol. 13959 (2023).
  \doi{10.1007/978-3-031-35504-2_3}

\bibitem{DBLP:conf/nana/WangLSMWYSLZD17}
Wang, Y., Liu, L., Su, C., Ma, J., et~al.: {CryptSQLite}: Protecting data
  confidentiality of {SQLite} with {Intel} {SGX}. In: {NaNA}'17 (2017).
  \doi{10.1109/NANA.2017.48}

\bibitem{DBLP:conf/middleware/YuhalaMFST0GL21}
Yuhala, P., M{\'{e}}n{\'{e}}trey, J., Felber, P., Schiavoni, V., et~al.:
  Montsalvat: {Intel} {SGX} shielding for {GraalVM} native images. In:
  Middleware '21. {ACM} (2021). \doi{10.1145/3464298.3493406}

\bibitem{Zhang24}
Zhang, R., Gerlach, L., Weber, D., Hetterich, L., et~al.: {CacheWarp}:
  Software-based fault injection using selective state reset. In: {USENIX}
  Security 2024 (2024)

\end{thebibliography}

\end{document}